\documentstyle[amssymb,aps]{revtex}

\begin{document}
\draft
\title{Two-dimensional small-world networks: navigation with local information}
\author{Jian-Zhen Chen$^{a,b}$ Wei Liu$^a$ and Jian-Yang Zhu$^a$\thanks{%
Author to whom correspondence should be addressed; Electronic address:
zhujy@bnu.edu.cn}}
\address{$^a$Department of Physics, Beijing Normal University, Beijing 100875, China\\
$^b$Department of Physics, JiangXi Normal University, Nanchang 330027, China}
\maketitle

\begin{abstract}
Navigation process is studied on a variant of the Watts-Strogatz
small world network model embedded on a square lattice. With
probability $p$, each vertex sends out a long range link, and the
probability of the other end of this link falling on a vertex at
lattice distance $r$ away decays as $ r^{-\alpha }$. Vertices on the
network have knowledge of only their nearest neighbors. In a
navigation process, messages are forwarded to a designated target.
For $\alpha <3$ and $\alpha \neq 2$, a scaling relation is found
between the average actual path length and $pL$, where $L$ is the
average length of the additional long range links. Given $pL>1$,
dynamic small world effect is observed, and the behavior of the
scaling function at large enough $pL$ is obtained. At $\alpha =2$
and $3$, this kind of scaling breaks down, and different functions
of the average actual path length are obtained. For $\alpha >3$, the
average actual path length is nearly linear with network size.

\end{abstract}


\section{Introduction}

The famous experiment of messages being forwarded to a target among a group
of people, carried out by Milgram \cite{Milgram} in the 1960s, and also by
Dodds {\it et al.} in 2003 \cite{Dodds} on a larger scale, reveals the
existence of short paths between pairs of distant vertices in networks that
appear to be regular (i.e. the small-world effect \cite
{Milgram,WS,Watts,effect1,effect2}). One of the important quantities that
characterize this small-world effect is the average shortest path length $%
\overline{\left\langle d\right\rangle }$ between two vertices. On small
world networks, this value grows very slowly (relative to the case of a
fully regular network) with the network size $N$. Recent empirical research
has shown that a great variety of natural and artificial networks with their
structure dominated by regularity are actually small worlds, and their
average path lengths grow as $\ln N$\cite{log}, or more slowly. (See Refs. 
\cite{review} for more reviews.)

An alternative issue revealed by experiments, but less obvious, is about the
realistic process of passing information on small world networks. This kind
of information navigation has been studied by Kleinberg\cite{Kleinberg}.
This process goes dynamically: When a message is to be sent to a designated
target, each individual forwards the message to one of its nearest neighbors
(connected either by a regular link or a shortcut) based on its limited
information. Without information of the whole network structure, this actual
path is usually longer than the shortest one given by the topological
structure. While $\overline{\left\langle d\right\rangle }$ is the average
shortest path length, the average actual path length $\overline{\left\langle
l\right\rangle }$ is the average number of steps required to pass messages
between randomly chosen vertex pairs. As $\overline{\left\langle
d\right\rangle }$ is usually referred to as the diameter of the system, in
the rest of this paper $\overline{\left\langle l\right\rangle }$ shall be
taken as the effective diameter, and $\overline{\left\langle d\right\rangle }%
\leq \overline{\left\langle l\right\rangle }$. It has been noted that the
topology of the network may significantly affect the behavior of $\overline{%
\left\langle l\right\rangle }$ \cite{Kleinberg,Watts,Zhu,Moura,topo}. In
other words, it may determine the efficiency of passing information.

Based on Milgram's experiment, Kleinberg studied navigation process on a
variant of the Watts-Strogatz (W-S) small-world model \cite{WS} on an $%
N\times N$ open regular square lattice. Each vertex sends out a long range
link with probability $p$, and the probability of the other end falling on a
vertex at Euclidean distance $r_{e}$ away decays as $r_{e}^{-\alpha }$.
Kleinberg studied $\overline{\left\langle l\right\rangle }$ when each vertex
sends out one long range link, and proved a lower bound of $\overline{%
\left\langle l_{p=1}\right\rangle }=cN^{\beta \left( \alpha \right) }$. When 
$\alpha =0$, the long range links are uniform, and $\overline{\left\langle
l_{\alpha =0,p=1}\right\rangle }\propto N^{2/3}$ was obtained. de Moura {\it %
et al. }\cite{Moura} studied $\overline{\left\langle l\right\rangle } $ on
the $D$-dimensional W-S model with $\alpha =0$ and varying $p$, and obtained 
$\overline{\left\langle l_{\alpha =0,p=1}\right\rangle }\propto N^{1/\left(
D+1\right) }$, and thus $\overline{\left\langle l_{\alpha
=0,p=1}\right\rangle }\propto N^{1/3}$ in the two-dimensional case. In the
more recent work of Zhu {\it et al. }\cite{Zhu} on the one-dimensional case,
the variance of $\overline{\left\langle l_{\alpha ,p}\right\rangle }$ with
both $\alpha $ and $p$ was studied, and scaling relations were shown to
exist. For the studies of the searching processes on other different
networks, see Refs. \cite{Watts,scalefree}.

In this paper, we systematically investigate the navigation process on a
two-dimensional variant of the W-S network model \cite{WS,chemi}. We study
the behavior of $\overline{\left\langle l\right\rangle }$ by first working
out the scaling relations in the two-dimensional case. Our result also
provides new understanding of the scaling analysis in Ref. \cite{Zhu}. In
Sec. \ref{Sec.2}, the model used here is constructed and the navigation
process is described, and then the average actual path length $\overline{%
\left\langle l\right\rangle }$ is obtained with some approximation based on
a rigorous treatment. Following that, in Sec. \ref{Sec.3}, the dependence of 
$\overline{\left\langle l\right\rangle }$ on $N,\alpha ,$ and $p$ are
presented based on scaling relations. Special attention is paid to the cases
studied in the works of Kleinberg \cite{Kleinberg} and de Moura \cite{Moura}%
. Summary and discussions can be found in Sec. \ref{Sec.4}.

\section{The construction of the navigation model}

\label{Sec.2}

Our model starts from a $N\times N$ two-dimensional square lattice. With
periodic boundary condition, the lattice distance between two vertices $%
(x,y) $ and $(x^{\prime },y^{\prime })$ can be written in a two-dimensional
fashion as 
\begin{equation}
r_{\left( x,y\right) ,\left( x^{\prime },y^{\prime }\right) }=\Delta
x+\Delta y,\ \ \Delta x=N/2-\left| \left| x-x^{\prime }\right| -N/2\right|
,\ \ \Delta y=N/2-\left| \left| y-y\prime \right| -N/2\right| .  \label{rij}
\end{equation}
This value is actually the length of the shortest path connecting these two
vertices through only regular links. To generate a small world, with
probability $p$ ($0\leq p\leq 1$) each vertex sends out an additional link
to another vertex (excluding its original nearest neighbors)\footnote{%
It should be noted, though, that we do not consider direction of links in
the following discussions.}. If this other vertex is selected at random,
then we are creating a small-world network with random shortcuts. Based on
realistic considerations (for example, people tend to be brought together by
similar interest, occupation, etc.), we shall add the shortcuts in a biased
manner \cite{Kleinberg,chemi}: If the shortcut starts from vertex $%
i(x_i,y_i) $, the probability that vertex $j(x_j,y_j)$ is selected as the
end depends on the lattice distance between them in the following way, 
\[
P\left( r_{\left( x,y\right) ,\left( x^{\prime },y^{\prime }\right) }\right)
=\frac 1Ar_{\left( x,y\right) ,\left( x^{\prime },y^{\prime }\right)
}^{-\alpha }, 
\]
where $\alpha $ is a positive exponent and 
\[
A=\sum_{(x^{^{\prime \prime }},y^{\prime \prime })\neq (x,y),(x\pm
1,y),(x,y\pm 1)}r_{\left( x,y\right) ,\left( x^{\prime \prime },y^{\prime
\prime }\right) }^{-\alpha } 
\]
is the normalization factor.

In the model described above, the navigation process can be simulated with
the so-called ``greedy''\ algorithm \cite{Kleinberg}: Without loss of
generality, suppose the target is vertex $(0,0)$. At each step, the current
message holder, vertex $n(x_n,{y}_n)$, passes the message through one of its
regular or long-range links. Based on its limited local information, this
link is believed to bring the message the closest to the target. Based on
this algorithm, the actual path length $\left\langle l(x_n,y_n)\right\rangle 
$ can be obtained after taking an ensemble average over all possible
realizations of the network (with a set of fixed parameters, $p$, $N$, $%
\alpha $, etc.).

In the simplest case, we suppose that each vertex has information of only
the vertices that can be reached within one step, and do the calculation as
the following: (1) If the current message holder is vertex $\left(
0,1\right) $, we simply have 
\begin{equation}
\left\langle l\left( 0,1\right) \right\rangle =1.  \label{ditui1}
\end{equation}
It is the same for the other nearest neighbors of the target $\left(
0,-1\right) $, $\left( 1,0\right) $, and $\left( -1,0\right) $. (2) There
are $8$ nodes with lattice distance $2$ from the target: $\left( 0,\pm
2\right) $, $\left( \pm 2,0\right) $, $\left( \pm 1,\pm 1\right) $, and $%
\left( \mp 1,\pm 1\right) $. If the current message holder is, for example, $%
\left( 0,2\right) $, then with probability 
\[
W_{\left( 0,2\right) \rightarrow \left( 0,0\right) }=1-\left( 1-p\frac{%
2^{-\alpha }}A\right) ^2, 
\]
it is directly linked to the target via one shortcut, which means the
message is sent directly to the target with this probability. On the other
hand, the probability that the message is forwarded along a regular bond is 
\[
W_{reg}=1-W_{\left( 0,2\right) \rightarrow \left( 0,0\right) }. 
\]
Thus 
\begin{equation}
\left\langle l\left( 0,2\right) \right\rangle =1\times W_{\left( 0,2\right)
\rightarrow \left( 0,0\right) }+2\times W_{reg}.  \label{ditui2}
\end{equation}
The calculation is the same for the other $7$ nodes mentioned above. (3) In
a general case, the message is held by vertex $(x,y)$. $W_{\left( x,y\right)
\rightarrow \left( x^{\prime },y^{\prime }\right) }$ denotes the probability
that the message is forwarded in the next step to a vertex $\left( x^{\prime
},y^{\prime }\right) $, which must be nearer to the target than $\left(
x,y\right) $ by at least lattice distance $2$. If the message holder is not
able to find a shortcut, the message will be forwarded along a regular link
with probability 
\begin{equation}
W_{reg}=1-\sum_{\left( x^{\prime },y^{\prime }\right) }W_{\left( x,y\right)
\rightarrow \left( x^{\prime },y^{\prime }\right) }.  \label{Wreg}
\end{equation}
For example, if the vertex $\left( 3,2\right) $ passes the message through a
regular link, it will randomly choose $(2,2)$ or $(3,1)$, which in the
following will be denoted by $\left( x_{reg},y_{reg}\right) $.

Now, with this set of probabilities $W$, we obtain, 
\begin{equation}
\left\langle l\left( x,y\right) \right\rangle =W_{\left( x,y\right)
\rightarrow \left( 0,0\right) }+\sum_{\left( x^{\prime },y^{\prime }\right)
}W_{\left( x,y\right) \rightarrow \left( x^{\prime },y^{\prime }\right) }
\left[ 1+\left\langle l\left( x^{\prime },y^{\prime }\right) \right\rangle
\right] +W_{reg}\left[ 1+\left\langle l\left( x_{reg},y_{reg}\right)
\right\rangle \right] .  \label{dituin}
\end{equation}
Considering that $p/A$ is a relatively small quantity, $W_{\left( x,y\right)
\rightarrow \left( x^{\prime },y^{\prime }\right) }$ can be expressed as 
\cite{Zhu} 
\begin{equation}
W_{\left( x,y\right) \rightarrow \left( x^{\prime },y^{\prime }\right) }=2p%
\frac{r_{\left( x,y\right) ,\left( x^{\prime },y^{\prime }\right) }^{-\alpha
}}A,  \label{Wnm}
\end{equation}
where we have used the fact that $r_{\left( x,y\right) ,\left( x^{\prime
},y^{\prime }\right) }=r_{\left( x^{\prime },y^{\prime }\right) ,\left(
x,y\right) }$. Then it is an easy task to obtain $W_{reg}$ from Eq. (\ref
{Wreg}).

Recall that we have the definition of the average shortest path length $%
\overline{\left\langle d\right\rangle }=\frac 1{N^2(N^2-1)}\sum_{i\neq
j}d_{ij}$, where $d_{ij\text{ }}$is the length of the shortest path between
vertices $i$ and $j$. By contrast, the average actual path length can be
defined as 
\begin{equation}
\overline{\left\langle l\right\rangle }=\frac 1{N^2\left( N^2-1\right) }%
\sum_{\left( x,y\right) \neq \left( 0,0\right) }\left\langle
l(x,y)\right\rangle .  \label{avel}
\end{equation}
Further, with vertex $\left( 0,0\right) $ being the target, we group the
other nodes according to their lattice distance from the target, and in the
following we shall also discuss the function $\overline{\left\langle l\left(
n\right) \right\rangle }$, which for each value of $n$ is obtained by
averaging all nodes $\left( x,y\right) $ with $r_{\left( x,y\right) ,\left(
0,0\right) }=n$.

\section{Features of the navigation process}

\label{Sec.3}

The average actual path length depends on multiple parameters. Here we take
into consideration varying $N$, $\alpha $, and $p$, but keep the range of
view of each vertex limited to its nearest neighbors. Our discussion of the
navigation process starts from looking for the basic scaling relations.

Scaling relation is not new in the theories of small world effect. Actually,
it plays a central role in the current theoretical framework. In 1999,
Newman \cite{Newman} showed that in the W-S model with uniform shortcuts the
average shortest path length is a function of $pN$, and it sharply decreases
when $pN$ becomes larger than 1. Newman noticed that $pN$ is simply the
expected number $M$ of long-range links. The threshold of small world
behavior is $M=pN>1$, that means the network becomes a small world when
there is more than one long-range link. When the model network is
generalized, this interpretation shall be generalized as well. For example,
in a discussion of the scaling relations in the problem of dynamic
navigation, Zhu {\it et al.} \cite{Zhu} considered inhomogeneous long range
links (the probability of linking two nodes falls when their lattice
distance increases). In their study, the dynamic small world behavior is
switched on when $ML^{\prime }>1$, where $M$ is the number of long-range
links and $L^{\prime }$ is the average reduced link length (the average
length of long range links $L$ divided by the system size). Although they
focused on different aspects (static and dynamic) of the small world effect,
we can still compare these two versions of scaling relations. Because in the
model studied by Newman, $L^{\prime }\sim 1/4$, it is consistent with the
interpretation of Zhu {\it et al}. Actually, as we shall see below, the
interpretation of Zhu {\it et al.} can be developed as well, when a more
general model is considered.

In the Introduction we have defined $\overline{\left\langle l\right\rangle }$
as the effective diameter, in the following we will use $\overline{%
\left\langle l^{\prime }\right\rangle }=\overline{\left\langle
l\right\rangle }/N$ as the reduced effective diameter. If the network is
regular, $\overline{\left\langle l^{\prime }\right\rangle }$ will appear as
a constant. Bearing this in mind, we first look at the results shown in
Fig.1. For each value of $\alpha $ (with exception at $\alpha =2$, as will
be shown below), $\overline{\left\langle l^{\prime }\right\rangle }$ appears
as a function of $pL$, where $L$ is the average length of long-range links.
Thus our study clearly supports an interpretation different from that in
Ref. \cite{Zhu}: Instead of $ML^{\prime }$, the parameter should be $pL$. In
the one-dimensional case, this equals $ML^{\prime }$, and thus consistent
with Ref. \cite{Zhu}. When $pL\ll 1$, $\overline{\left\langle l^{\prime
}\right\rangle }\rightarrow 0.5$, and $\overline{\left\langle l\right\rangle 
}\varpropto N$, indicating that the network is virtually regular, and when $%
pL$ increases beyond $1$, the system begins to show a dynamic small world
behavior.

However, we find interesting exceptions at $\alpha =2$ and $\alpha =3$. As
shown in Fig.2, at $\alpha =2$, $\exp \left( \overline{\left\langle
l^{\prime }\right\rangle }\times pL\right) $ is a linear function of $pL/\ln
N$ for $pL$ significantly larger than $1$. Due to this extra factor of $%
1/\ln N$, there is no way that $\overline{\left\langle l^{\prime
}\right\rangle }$ can be written as a function of $pL$ for $\alpha =2.$
Because at $\alpha =2$ the system shows the shortest $\overline{\left\langle
l\right\rangle }$, this is certainly a case of special interest. It becomes
more curious when we notice that in one dimension there is not such an
exception in the scaling analysis \cite{Zhu}. As shown in Fig.3, at $\alpha
=3$, $\overline{\left\langle l^{\prime }\right\rangle }$ appears as a
function of $pL\ln N$, and the dynamic small world behavior is seen when
this parameter exceeds $1$. There is more discussion of these interesting
points later.

Figs.1, 2 and 3 can give us more information once we get an idea of $L$. For 
$N$ large enough, we can use the following approximation,

\begin{equation}
L=\frac{\int_0^{N/2-1}dx\int_1^{N/2}(x+y)^{1-\alpha }dy}{\int_0^{N/2-1}dx%
\int_1^{N/2}(x+y)^{-\alpha }dy},  \label{L'}
\end{equation}
which gives 
\begin{equation}
L=\frac{1-\alpha }{3-\alpha }\frac{(N-1)^{3-\alpha }-2(N/2)^{3-\alpha }+1}{%
(N-1)^{2-\alpha }-2(N/2)^{2-\alpha }+1},~\left( \alpha \neq 0,1,2,3\right) ,
\label{ravel}
\end{equation}
\begin{equation}
L_{\alpha =0}=\frac 16\frac{N^3-2\left( \frac N2\right) ^3+1}{\left( \frac 12%
N-1\right) ^2},  \label{ravel0}
\end{equation}
\begin{equation}
L_{\alpha =1}=\frac{\left( \frac 12N-1\right) ^2}{\left( N-1\right) \ln
\left( N-1\right) -N\ln \left( \frac N2\right) },  \label{ravel1}
\end{equation}
\begin{equation}
L_{\alpha =2}=\frac{\left( N-1\right) \ln \left( N-1\right) -N\ln \left( 
\frac N2\right) }{2\ln \frac N2-\ln \left( N-1\right) },  \label{ravel2}
\end{equation}
and 
\begin{equation}
L_{\alpha =3}=\frac{2\ln \frac N2-\ln \left( N-1\right) }{\frac 1{2\left(
N-1\right) }-\frac 2N+\frac 12}.  \label{ravel3}
\end{equation}
Some comments on this approximation: We have chosen the upper and lower
limits for the integrals with some arbitrariness. For $\alpha \leqslant 3$
and for $N$ very large, it gives results good enough for the later
discussion. Surely this approximation fails for $\alpha >3$, but in that
region what is important is $L$ stays finite as $N$ goes to infinity.

From Eqs. (\ref{ravel}) to (\ref{ravel3}), we have for large enough $N$

\begin{equation}
L_{\alpha =0}\rightarrow \frac{N}{2},  \label{La0}
\end{equation}

\begin{equation}
L_{\alpha =1}\rightarrow \frac{N}{4\left( \ln 2\right) },  \label{La1}
\end{equation}

\begin{equation}
L=\frac{1-\alpha }{3-\alpha }\cdot \frac{1-2^{\alpha -2}}{1-2^{\alpha -1}}%
N,~\left( 0<\alpha <1,~1<\alpha <2\right) ,  \label{La02}
\end{equation}

\begin{equation}
L_{\alpha =2}\rightarrow \ln 2\frac{N}{\ln N},  \label{La2}
\end{equation}

\begin{equation}
L\rightarrow N^{3-\alpha }\frac{1-\alpha }{3-\alpha }\cdot \left(
1-2^{\alpha -2}\right) \rightarrow 0,~\left( 2<\alpha <3\right)  \label{La23}
\end{equation}

\begin{equation}
L_{\alpha =3}\rightarrow 2\ln N,  \label{La3}
\end{equation}

\begin{equation}
L\rightarrow \text{finite},~\left( \alpha >3\right) .  \label{La34}
\end{equation}

In the above equations, $\alpha =0,2,3$ come out as special points. $\alpha
=0$, corresponds to the totally random network. Below $\alpha _c^1=D=2$, $%
L_{0\leq \alpha <2}$ is always proportional to $N$. In Ref. \cite{Sen}, the
authors proposed that for $0\leq \alpha <\alpha _c^1$ the system is in the
random network phase, and $\alpha _c^1$ is the continuous phase transition
point from the random network phase to the small-world phase. As $\alpha $
increases above $\alpha _c^2=D+1=3$, $L$ is finite and independent of $N$ in
the limit of $N\rightarrow \infty $. As a result, the system is virtually a
regular network for $\alpha >\alpha _c^2$.

Now we have the scaling relations shown in Figs.1, 2 and 3, and $L$ as a
function of $N$ given by Eqs. (\ref{La0}) to (\ref{La34}). In the following
we shall discuss the system behavior with $\alpha $ starting from zero.

(1) $0\leq \alpha <\alpha _c^1$: As shown in Fig.1, when $pL\ll 1$, $%
\overline{\left\langle l^{\prime }\right\rangle }\rightarrow 0.5$,
indicating that the network is virtually regular. When $pL$ increases beyond 
$1$, the system begins to show a dynamic small world behavior, in the sense
that 
\begin{equation}
\overline{\left\langle l^{\prime }\right\rangle }\sim \left( pL\right)
^{-\gamma }\sim p^{-\gamma }N^{-\gamma },
\end{equation}
or 
\begin{equation}
\overline{\left\langle l\right\rangle }\sim p^{-\gamma }N^{1-\gamma },
\end{equation}
where $\gamma $ depends only on $\alpha $. From the linear fit, at $\alpha
=0 $ we have obtained that $\gamma \approx 0.309$ and $\overline{%
\left\langle l\right\rangle }\sim p^{-0.309}N^{0.691}$. Note that $\gamma $
is obtained from linear fit of limited data and cannot be exact, but our
result, $\beta =1-\gamma \approx 0.691$, is close to Kleinberg's result, $%
\beta =2/3$. (We observe that in Ref. \cite{Moura}, $\beta =1/3$.) As $%
\alpha $ increases, $\gamma $ increases, but $\beta $ remains positive as $%
\alpha \rightarrow \alpha _c^1$. With our present results, we are not able
to give the full function of $\gamma \left( \alpha \right) $, because near $%
\alpha =2$, $\overline{\left\langle l\right\rangle }$ as a function of $N$
severely deviates from a power law for relatively small $N$.

(2) $\alpha =\alpha _c^1=2$: At this point, when $pL/\ln N\ll 1$, $\overline{%
\left\langle l^{\prime }\right\rangle }\rightarrow 0.5$, showing that it is
a regular network. When $pL/\ln N\ $increases above $1$, $\exp \left( 
\overline{\left\langle l^{\prime }\right\rangle }\times pL\right) $ turns
into a linear function of $pL/\ln N$, and it means 
\begin{equation}
\overline{\left\langle l^{\prime }\right\rangle }\sim \ln \left( pL/\ln
N\right) /pL.
\end{equation}
With $L\sim \ln 2\frac N{\ln N}$ for $N\rightarrow \infty $, this gives 
\begin{equation}
\overline{\left\langle l\right\rangle }\sim \frac{\ln N\left[ \ln p+\ln
\left( \ln 2\right) +\ln N-2\ln \left( \ln N\right) \right] }{p\ln 2},
\end{equation}
or, in the limit of $N\rightarrow \infty ,$ 
\begin{equation}
\overline{\left\langle l\right\rangle }\sim \frac{\ln N\left( \ln p+\ln
N\right) }p.
\end{equation}
We have studied the case of $0\leq \alpha <\alpha _c^1$, and the case of $%
\alpha >\alpha _c^1$ will be studied below. We shall see that at $\alpha
=\alpha _c^1$ the smallest $\overline{\left\langle l\right\rangle }$ is
achieved.

(3) $\alpha _c^1<\alpha <\alpha _c^2$: Regularity dominates for $pL\ll 1$.
For $pL>1$, dynamic small world effect arises. Similar to the case of $0\leq
\alpha <\alpha _c^1$, we have once again obtained 
\begin{equation}
\overline{\left\langle l\right\rangle }\sim p^{-\gamma }N^{1-\gamma },
\end{equation}
and $\gamma $ tends to zero as $\alpha $ approaches $\alpha _c^2$.

(4) $\alpha =\alpha _c^2=3$: When $pL\ln N\ll 1$, $\overline{\left\langle
l^{\prime }\right\rangle }\rightarrow 0.5$, the network also shows
dominating regularity. When $pL\ln N$ greatly exceeds $1$, 
\begin{equation}
\overline{\left\langle l^{\prime }\right\rangle }\sim \left( pL\ln N\right)
^{-\gamma }.
\end{equation}
Substitute $L\sim 2\ln N$ into the above equation, then for large enough $N$%
, we have 
\begin{equation}
\overline{\left\langle l\right\rangle }\sim p^{-\gamma }N\left( \ln N\right)
^{-2\gamma },
\end{equation}
but the size of the networks used in the present study prevented us from
getting a accurate estimate of $\gamma $.

(5) $\alpha >\alpha _c^2$: Since in this case $L$ stays finite when $%
N\rightarrow \infty $, the system is believed to behave like a regular
network. It is confirmed by the numerical calculation, which gives $%
\overline{\left\langle l\right\rangle }$ nearly linear as $N$.

\section{Summary and discussion}

\label{Sec.4}

In this work, we investigate the navigation process on a variant of
Watts-Strogatz (W-S) model embedded on a two-dimensional square lattice with
periodic boundary condition. With probability $p$, each vertex sends out a
long range link, and the probability that the other end of this link falls
on a vertex at lattice distance $r$ away decays as $r^{-\alpha }$. Vertices
on the network have knowledge of only their nearest neighbors. In a
navigation process, messages are forwarded to a designated target, and the
average actual path length $\overline{\left\langle l\right\rangle }$ is
obtained with varying $\alpha $, $p$, and $N$.

Our result is consistent with the existence of two phase transitions at $%
\alpha _c^1=D=2$ (random network to small world network), and $\alpha
_c^2=D+1=3$ (small world network to regular network). For $\alpha <\alpha
_c^2$, and $\alpha \neq \alpha _c^1$, it is found that $\overline{%
\left\langle l^{\prime }\right\rangle }\equiv \overline{\left\langle
l\right\rangle }/N\sim f_\alpha \left( pL\right) $, where $L$ is the average
length of the additional long range links. This develops the scaling
analysis in the works of Newman \cite{Newman} and Zhu {\it et al. }\cite{Zhu}%
. Given $pL>1$, dynamic small world effect is observed, and the behavior of $%
f_\alpha $ at large enough $pL$ gives $\overline{\left\langle l\right\rangle 
}\sim p^{-\gamma \left( \alpha \right) }N^{1-\gamma \left( \alpha \right) }$%
. When $\alpha =0$, $\gamma $ is close to $1/3$, so $\beta =1-\gamma \sim
2/3 $, in agreement with Kleinberg's result of $\overline{\left\langle
l\right\rangle }\varpropto N^{2/3}$ for $p=1$. As $\alpha \ $increases, $%
\gamma $ increases (but stays below $1$), and once $\alpha $ exceeds $\alpha
_c^1$, $\gamma $ begins to decrease, and $\gamma $ approaches zero as $%
\alpha \rightarrow \alpha _c^2$. At $\alpha =\alpha _c^1$, this kind of
scaling breaks down, and $\overline{\left\langle l^{\prime }\right\rangle }$
can no longer be written as a function of $pL$. In this case we can still
get $\overline{\left\langle l\right\rangle }\sim \ln N\left( \ln p+\ln
N\right) /p$ for $N$ large enough. Note that only at $\alpha =\alpha _c^1$, $%
\overline{\left\langle l\right\rangle }$ grows as a polynomial of $\ln N$,
and it is the closest point to the static small world effect \cite{Zhu}. At $%
\alpha =\alpha _c^2$, the scaling is $\overline{\left\langle l^{\prime
}\right\rangle }\sim f\left( pL\ln N\right) $, and accordingly $\overline{%
\left\langle l\right\rangle }\sim p^{-\gamma }N\left( \ln N\right)
^{^{-2\gamma }}$ with large enough $pL\ln N$. For $\alpha >\alpha _c^2$, $%
\overline{\left\langle l\right\rangle }$ is nearly linear with $N$.

It is reasonable that in social networks (like various other networks) the
probability of connection falls as distance (in various senses, e.g.,
occupation) increases, and the apparently very small value of $\overline{%
\left\langle l\right\rangle }$ in human society \cite{Milgram,Dodds}
suggests human society might have its exponent $\alpha $ being close to $%
\alpha _{c}^{1}$.

The great success of the idea of small world has since motivated much effort
in studying various dynamic processes based on the small world network
model. The limited knowledge of the nodes of a network is an important
limitation that has to be considered when studying navigation processes.
Another interesting and important limitation is due to the fact that the
links in a network are usually associated with "weights", as systematically
studied in Refs. \cite{weight}. Further studies on link-weighted small-world
model should help us gain insight in the navigation and other relevant
phenomena in various artificial and natural networks, and help us design
networks with higher efficiency.

\acknowledgments

The work was supported by the National Natural Science Foundation of China
(No. 10375008), and the National Basic Research Program of China
(2003CB716302). We thank H. Zhu for helpful discussions.

\centerline{\bf Caption of figures} \vskip1cm

Fig.1. (Color online) The reduced average actual path length $\overline{%
\left\langle l^{\prime }\right\rangle }$ varies as $pL$ for $\alpha =0,1,2.5$%
, where $L$ is the average length of the additional long range links. The
data collapse with each specific $\alpha $ contains curves with $%
N=200,400,600,800,1000$ respectively. On each curve with fixed $N$, $%
p=1.3^{-i}$, where $i=0,1,2,...,47$ ($p$ has the same set of values on each
curve in Fig.2 and Fig.3). The solid line $y\sim x^{-0.309}$ is a guide to
the eye.

Fig.2. (Color online) With given $\alpha =2$, plots show the relation
between $\exp \left( \overline{\left\langle l^{\prime }\right\rangle }\cdot
pL\right) $ and $pL$ at $N=200,400,600,800,1000$.

Fig.3. (Color online) The relationship between $\overline{\left\langle
l^{\prime }\right\rangle }$ and $pL\ln N$ in log-log scale at $\alpha =3$
for $N=200,400,600,800,1000$.

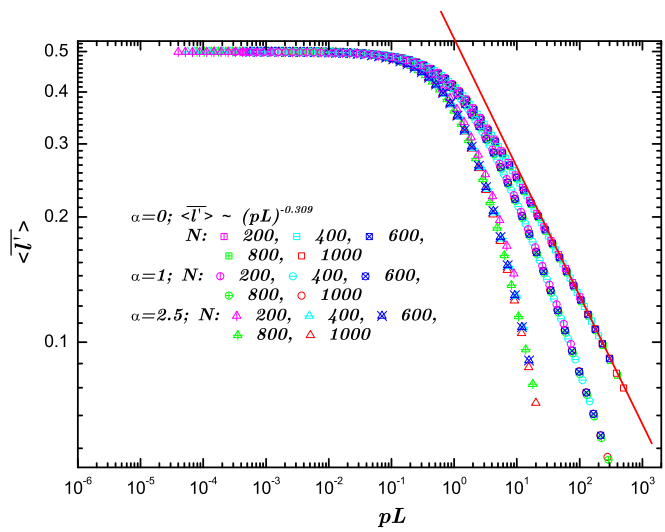
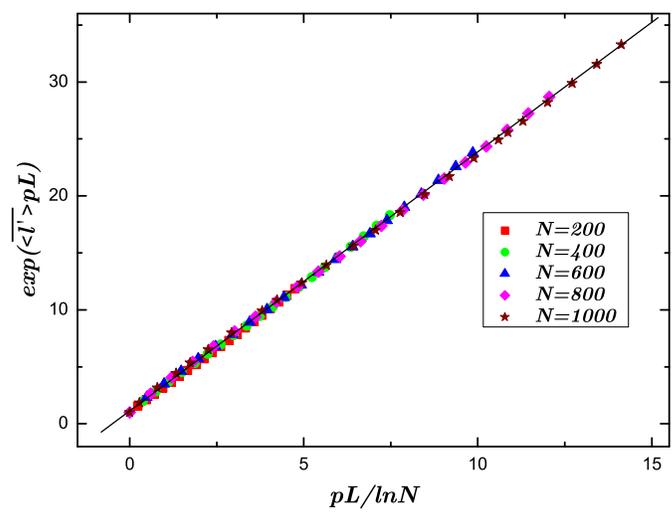
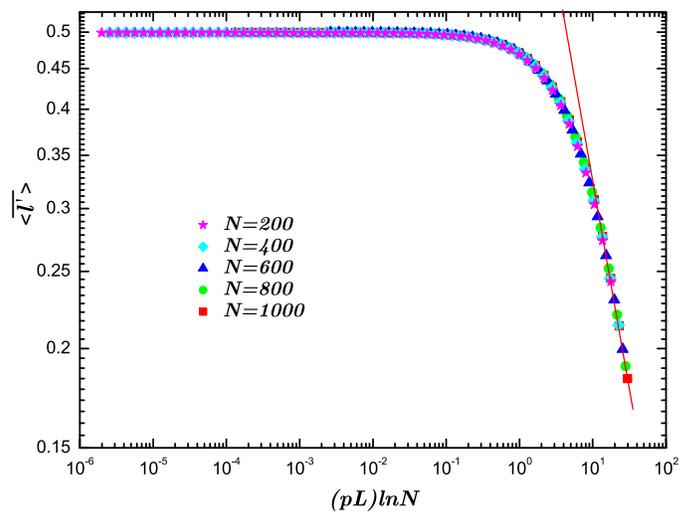
\end{document}